\documentclass[aps,prb,nobibnotes,preprint,amsmath,amssymb,superscriptaddress,floatfix]{revtex4-1}
\usepackage{graphicx, multirow, xcolor, bm, ulem}
\usepackage{amsfonts,amssymb,amsmath}
\usepackage{graphicx,dcolumn,bm,color,braket,slashed}
\usepackage{times, comment, mathtools, extarrows,textcomp,ulem}

\begin{document}

\title{
Revealing inverted chirality of hidden domain wall states in multiband systems without topological transition
}

\author{Seung-Gyo Jeong }
\thanks{These authors contributed equally to this work.}
\affiliation{Department of Physics, Pohang University of Science and Technology, Pohang 37673, South Korea}

\author{Sang-Hoon Han}
\thanks{These authors contributed equally to this work.}
\affiliation{Department of Physics, Hanyang University, Seoul 04763, South Korea}

\author{Tae-Hwan Kim}
\email{taehwan@postech.ac.kr}
\affiliation{Department of Physics, Pohang University of Science and Technology, Pohang 37673, South Korea}

\author{Sangmo Cheon}
\email{sangmocheon@hanyang.ac.kr}
\affiliation{Department of Physics, Hanyang University, Seoul 04763, South Korea}
\affiliation{Research Institute for Natural Science and High Pressure, Hanyang University, Seoul 04763, South Korea}

\begin{abstract}
Chirality, a fundamental concept from biological molecules to advanced materials, is prevalent in nature. Yet, its intricate behavior in specific topological systems remains poorly understood. Here, we investigate the emergence of hidden chiral domain wall states using a double-chain Su-Schrieffer-Heeger model with interchain coupling specifically designed to break chiral symmetry. Our phase diagram reveals single-gap and double-gap phases based on electronic structure, where transitions occur without topological phase changes. In the single-gap phase, we reproduce chiral domain wall states, akin to chiral solitons in the double-chain model, where chirality is encoded in the spectrum and topological charge pumping. In the double-gap phase, we identify hidden chiral domain wall states exhibiting opposite chirality to the domain wall states in the single-gap phase, where the opposite chirality is confirmed through spectrum inversion and charge pumping as the corresponding domain wall slowly moves. By engineering gap structures, we demonstrate control over hidden chiral domain states. Our findings open avenues to investigate novel topological systems with broken chiral symmetry and potential applications in diverse systems.
\end{abstract}

\maketitle

\clearpage

\section*{Introduction}
Chirality and topology are concepts of great importance that lead to novel physical properties and potential applications in various fields.
Examples include topological surface states in topological insulators~\cite{hasan2010, qi2011}, Majorana fermions in topological superconductors~\cite{kitaev2001, elliott2015}, chiral stacking orders in charge density waves~\cite{xu2020, jiang2021, kim2022}, and topological lasers in photonic systems~\cite{st2017, bandres2018}.
As prototypical systems, the Su-Schrieffer-Heeger (SSH)~\cite{SSH1979} and Rice-Mele~\cite{RM1982} models exhibit exotic topological properties such as highly robust Jackiw-Rebbi domain wall zero-energy states~\cite{jackiw1976}, charge fractionalization~\cite{goldstone1981}, and spin-charge separation~\cite{jackiw1981}.
As a coupled SSH model, the double-chain (DC) model with broken chiral symmetry shows chiral solitons having topological chiral degrees of freedom and $Z_4$ topological algebraic operation~\cite{han2020}, where the chirality manifests as a spectrum of the chiral soliton and topological charge pumping observed during the adiabatic process as the chiral soliton slowly moving.

Such SSH and Rice-Mele models have been experimentally realized in various physical systems---polyacetylene~\cite{SSH1979, heeger1988}, cold atomic systems~\cite{atala2013, cooper2019}, artificial electronic lattices~\cite{drost2017, huda2020}, photonic systems~\cite{meier2016, ozawa2019}, and acoustic systems~\cite{zhou2017, zeng2021}.
Their quantized Berry phases~\cite{berry1984} are shown to be consistent with the bulk-boundary correspondence~\cite{hatsugai1993, bernevig2013topological}.
While many coupled SSH chain systems with nontrivial topology have been extensively studied~\cite{arkinstall2017, Zurita2021, Luo2022, Matveeva2023}, most possess chiral symmetry, precluding the emergence of chirality as chirality necessitates symmetry breaking.
In contrast, the DC model with broken chiral symmetry has been demonstrated in limited physical systems such as self-assembled indium nanowires and artificial atomic chains, exhibiting distinct chiral domain wall states~\cite{cheon2015, huda2020, jeong2022}. Despite being in the same topological class with preserved time-reversal and broken chiral symmetries~\cite{han2020, oh2021, schnyder2008, chiu2016}, a comprehensive understanding of such multiband systems remains elusive. Therefore, this work endeavors to present a unified framework elucidating the chirality, topology, and bulk-boundary correspondence underlying the emergence of chiral domain wall states in the coupled SSH chain systems with broken chiral symmetry.

Utilizing a representative DC model with interchain coupling whose chiral symmetry is broken, we unveil the emergence of hidden chiral domain wall states possessing inverted chirality even without necessitating any topological phase transition.
Our investigation yields a phase diagram revealing single- and double-gap phases depending on the dimerization of each SSH chain and the strength of the interchain coupling. Within the single-gap phase, the chiral domain wall states manifest two localized states akin to chiral solitons observed in the DC model~\cite{cheon2015}. In the double-gap phase, we observe the emergence of hidden chiral domain wall states characterized by opposite chirality compared to the preexisting domain wall states in the single-gap phase.

We physically verify this opposite chirality through spectrum inversion of the domain wall state and counter-directional charge pumping observed during the adiabatic process as the domain wall state slowly moves.
Using the extended two-dimensional effective Hamiltonian corresponding to the adiabatic process and the Berry curvature distribution,
we topologically confirm the chirality of hidden chiral domain wall states. 
Furthermore, by engineering the gap structure via tuning of the interchain coupling, we successfully control the emergence of the hidden chiral domain state.
Our results not only provide insights into the fundamental physics of multiband SSH systems with broken chiral symmetry but also have important implications for the design of novel devices based on chiral domain wall states.

\section*{Results and discussion}

\begin{figure}[t]
\centering
\includegraphics[width=.7\linewidth]{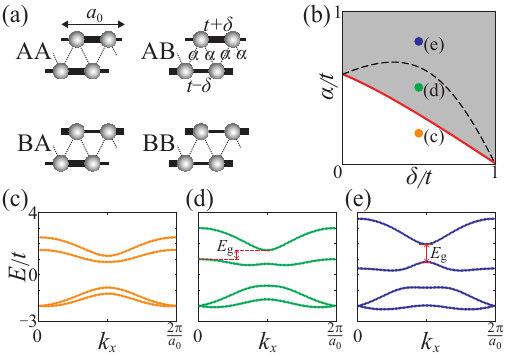}
\caption{\label{fig.1.gapphase} 
\textbf{Double-chain model having single and double gaps.}
\textbf{a} 
Double-chain (DC) model and its
geometric configurations for the four dimerized states, which are denoted as AA, AB, BA, and BB states.
Gray circles represent atoms with a single $p$-orbital with $t>0$.
In our model, $a_0$ represents the size of the unit cell.
Like the Su-Schrieffer-Heeger (SSH) model, the nearest electron hopping amplitude in the horizontal direction appears alternately with $t+\delta$ and $t-\delta$ due to the Peierls distortion.
$\alpha$ indicates the interchain hopping amplitude between lower and upper SSH chains.
\textbf{b} Phase map with respect to $\delta$ and $\alpha$. 
The red line denotes a phase boundary between single- and double-gap phases.
The double-gap phase region is divided by the black dashed line.
Above (below) the black dashed line, the upper gap is a direct (indirect) gap.
\textbf{c--e} Representative band structures of the DC models for \textbf{c} the single-gap phase, 
\textbf{d} the double-gap phase with the indirect upper gap, and \textbf{e} the double-gap phase with the direct upper gap, where $E/t$ indicates the dimensionless energy and $E_{\rm g}$ is the energy gap.
The parameter sets ($\alpha/t$, $\delta/t$) are given by (0.2, 0.5), (0.5, 0.5), and (0.8, 0.5) for \textbf{c}, \textbf{d}, and \textbf{e}, respectively.
Data in \textbf{b}--\textbf{e} are obtained from the AA groundstate.
Due to the degeneracy, the band structures and phase diagram are identical for the four types of configurations.
}
\end{figure}

\textbf{Double-chain model.} 
First, we introduce the DC model consisting of two SSH chains with interchain coupling (Fig.~\ref{fig.1.gapphase}a).
The interchain coupling acts as a tuning parameter that controls the electronic structure of this model while it was treated as a small perturbation in the previous works~\cite{cheon2015, han2020}.
Combining two SSH Hamiltonians, we get the Hamiltonian of the DC model:
\begin{eqnarray}
H_\text{DC} &=& H_\text{SSH}^{(1)}+H_\text{SSH}^{(2)}+H_\text{coupling},
\label{eq-2-DC-Hamiltonian}
\\
H_\text{SSH}^{(i)} &=& \sum_{n} t^{(i)}_{n+1,n} \, c^{(i)\dagger}_{n+1} c^{(i)}_{n}  + h.c. , \nonumber \\
H_\text{coupling} &=& \alpha (c_{n}^{(1)\dagger} c_{n}^{(2)} + c_{n}^{(1)\dagger} c_{n+1}^{(2)} + h.c. ), \nonumber
\end{eqnarray}
where  spin is abbreviated.
The superscript $(i=1,2)$ represents the upper and lower chains.
$c^{(i)\dagger}_{n}$ $(c_{n}^{(i)})$ denotes a creation (annihilation) operator for the $n$th site of the $i$th chain.
$t^{(i)}_{n+1,n} = t + (-1)^{n+1} \Delta^{(i)}$ indicates the horizontal nearest hopping integral for the $i$th chain,
where $t$ ($>0$) and $\Delta^{(i)}$ represent a hopping amplitude in the absence of the dimerization and the energy-valued dimerization displacement of the $i$-th chain, respectively.
$\alpha$ denotes the interchain coupling strength between the lower and upper SSH chains.
Since the A and B dimerized states are two degenerate groundstates for each SSH chain, the DC model naturally leads to four degenerate groundstates~\cite{cheon2015}, which are denoted as AA, AB, BA, and BB states (Fig.~\ref{fig.1.gapphase}a).
For instance, the AA groundstate is characterized by $\Delta^{(1)}=\Delta^{(2)}=\delta>0$, while the BB groundstate exhibits $\Delta^{(1)}=\Delta^{(2)}= - \delta <0$.
The DC model is classified into the AI class due to the broken chiral symmetry~\cite{cheon2015, han2020}, while the SSH model belongs to the BDI class due to the preserved time-reversal and chiral symmetries.
Such chiral symmetry breaking of the DC model provides the realization of the chirality of the nontrivial domain wall states~\cite{cheon2015, han2020}.

Figure~\ref{fig.1.gapphase}b--e shows the calculated phase map as a function of $\alpha/t$ and $\delta/t$ and three representative electronic band structures.
Depending on the number of gaps, there are two large distinct regions:
single- and double-gap phases (Fig.~\ref{fig.1.gapphase}b).
Figure~\ref{fig.1.gapphase}c shows a single gap between the second and third bands from the bottom, while Fig.~\ref{fig.1.gapphase}d,e has an additional gap between the third and fourth bands.
Furthermore, the region of the double-gap phase is divided into two subregions depending on whether the additional gap is direct or indirect: the upper gap between the red solid and black dashed curves is indirect (Fig.~\ref{fig.1.gapphase}d), while the upper gap becomes direct above the black dashed curve (Fig.~\ref{fig.1.gapphase}e).

In the SSH model, a one-dimensional one-band metallic chain at half filling undergoes the Peierls dimerization, which results in a two-band topological insulator~\cite{SSH1980, SSH1988, hasan2010}.
Similarly, the DC model becomes a four-band insulator after the dimerization~\cite{cheon2015, han2020}, which leads to the gap opening between the second and third bands near the Fermi level.
On the other hand, the gap-opening mechanism between the third and fourth bands is different due to the strong interchain coupling.
As the interchain coupling increases, the energy eigenvalue at $k_x=0$ of the third band decreases while the energy eigenvalue at $k_x=\pi/a_0$ ($a_0$ is the unit cell size) of the fourth band increases (Fig.~\ref{fig.1.gapphase}c--e). 
Such behavior eventually generates another gap between the third and fourth bands when the interchain coupling is larger than the phase boundary (red line in Fig.~\ref{fig.1.gapphase}b).
Analytically, the phase boundary between single- and double-gap phases is given by $\alpha_{1} = 2t+\delta - \sqrt{2 t^{2} +4 t \delta + 3 \delta^{2}}$.
Moreover, the system undergoes an indirect-direct gap transition with increasing interchain coupling.
The indirect-direct gap transition boundary (dashed line in Fig.~\ref{fig.1.gapphase}b) reads as $\alpha_{2} = 2t-\delta-\sqrt{2t^{2}-4 t \delta + 3 \delta^{2}}$.  
Additionally, the inversion symmetry protects the gap-closing between the first and second bands at the Brillouin zone boundary regardless of interchain coupling, the details of which are provided in Supplementary Note 1.

\begin{figure}[t]
\centering
\includegraphics[width=\linewidth]{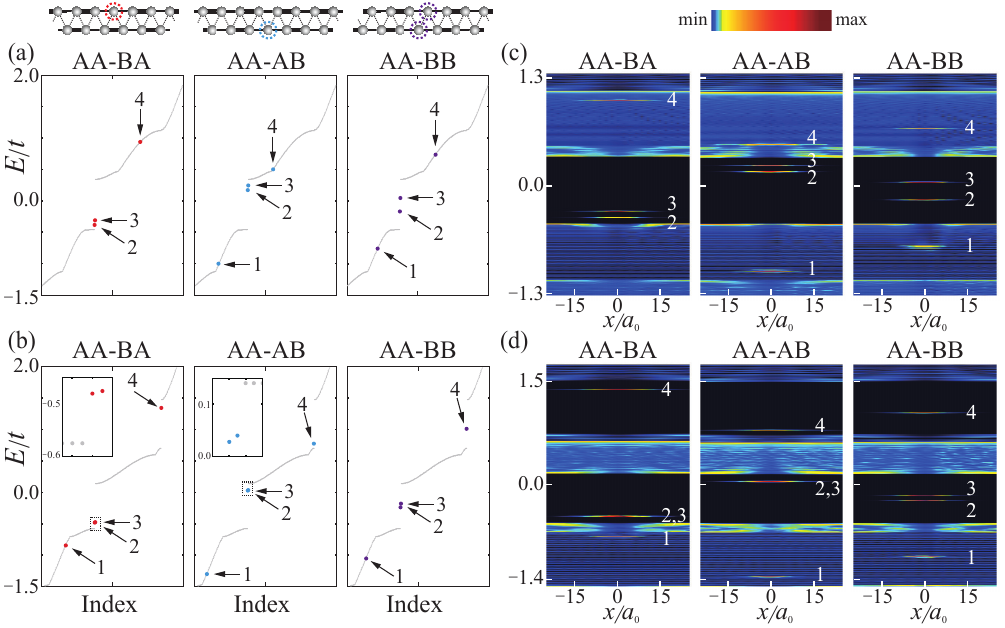}
\caption{\label{fig.2.domain_wall}
\textbf{Spectra and local density of state of localized domain wall states.} 
\textbf{a} \textbf{(b)} Electronic spectra and \textbf{c} \textbf{(d)} local density of state (LDOS) of the AA-BA, AA-AB, and AA-BB domain wall states in the single-gap (double-gap) phase. 
Red, blue, and purple dots denote the AA-BA, AA-AB, and AA-BB domain wall states, respectively, while gray dots indicate the bulk spectra.
In the single-gap phase, states 2 and 3 represent the in-gap domain-wall states, while states 1 and 4 do the hidden ones.
The state 1 in \textbf{a} and \textbf{c} is hard to distinguish, as discussed in the main text.
In the double-gap phase, states 2--4 represent the in-gap domain-wall states, while state 1 does the hidden one.
The LDOS plots show that all domain wall states are confined near the centers of the domain walls.
The parameter sets ($\alpha/t$, $\delta/t$) are given by (0.44, 0.3) and (0.7, 0.3) for \textbf{a,c} and \textbf{b,d}, respectively.
The inset above \textbf{a} shows the geometric configurations for AA-BA, AA-AB, and AA-BB domain wall states.
The insets in \textbf{b} show the close-up of the states 2 and 3.
In \textbf{c,d}, the color bar indicates the intensity of the LDOS.
}
\end{figure}

\textbf{Geometric configurations and quantum spectra of chiral domain walls.}
We now discuss geometric configurations and quantum energy spectra for all possible domain wall states connecting different groundstates.
When two of four groundstates are connected, we find only three distinct types of geometric configurations for nontrivial domain wall states (inset in Fig.~\ref{fig.2.domain_wall}a) due to the equivalence between the same geometric configuration of domain walls~\cite{cheon2015, han2020}. 
To distinguish such nontrivial geometries, we introduce the chirality and denote the AA-BA and AA-AB type configurations as right-chiral (RC) and left-chiral (LC) domain walls and the AA-BB type configuration as an achiral (AC) domain wall, following the notation of the previous works~\cite{cheon2015, han2020}.
For such three geometric configurations of nontrivial domain walls, we obtain the energy spectra and local density of states (LDOS) using tight-binding methods for single- and double-gap phases (Fig.~\ref{fig.2.domain_wall}).

Even though the same geometric domain wall configurations are employed for single- and double-gap phases, the electronic features of the localized domain wall states are quite different from each other.
For a single-gap phase, only two in-gap states (denoted as `2' and `3' in Fig.~\ref{fig.2.domain_wall}a,c) exist as localized domain wall states for all three types of domain walls. 
Two in-gap states for the AA-BA (AA-AB) geometric configuration are located below (above) the midgap, while two in-gap states for the AA-BB geometric configuration are located symmetrically with respect to the midgap.
The nontrivial positioning of the localized electronic states of the domain wall results from chiral symmetry breaking~\cite{cheon2015, kim2017}. This chiral symmetry breaking confers chirality upon the domain wall states, with chirality defined by the spectrum and topological charge pumping. These findings shed light on the interplay between chiral symmetry breaking and localized domain wall states, further enriching our understanding of the electronic behavior within such systems.

For a double-gap phase (Fig.~\ref{fig.2.domain_wall}b,d), an otherwise hidden in-gap state (denoted as `4')  emerges in the upper gap in addition to two in-gap states in the lower gap (denoted as `2' and `3').
The two in-gap states of each domain wall in the lower gap appear similar to those of the single-gap phase. 
Surprisingly, the in-gap state in the upper gap is located oppositely to those in the lower gap for the right- and left-chiral domain walls.
To clearly indicate such spectrum inversion of domain wall states between upper and lower gaps with respect to each midgap, we adopt the term `chirality inversion'.
The chiral inversion also occurs in the achiral AA-BB domain wall
even though the in-gap states for the lower and upper gaps seem not to be inverted due to the symmetrical positioning of in-gap states with respect to each midgap. 
Note that the topological meaning of the chirality inversion is the counter-directional charge pumping observed during the adiabatic process as the domain wall state slowly moves, which will be discussed in the next subsection.

\begin{figure}[t]
\centering
\includegraphics[width=.7\linewidth]{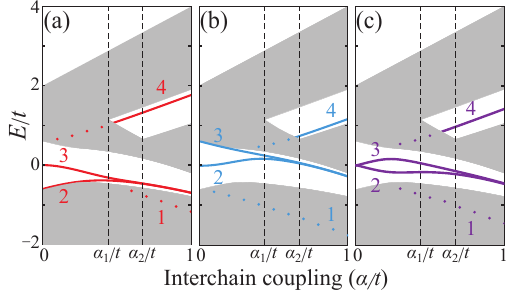}
\caption{\label{fig.3.evolution3}
\textbf{Evolution of energy spectra of domain wall states with respect to the interchain coupling.} 
Energy spectra for \textbf{a} AA-BA, \textbf{b} AA-AB, and \textbf{c} AA-BB domain walls.
Solid and dotted lines represent the in-gap domain wall states and hidden domain wall states, respectively.
Two domain wall states exist in the lower gap regardless of $\alpha/t$.
When the upper gap opens, a hidden domain wall state appears for all cases.
Indirect gaps exist between dashed lines ($\alpha_1 < \alpha < \alpha_2$) while direct gaps exist when $\alpha \ge \alpha_2$.
Here, $\delta /t= 0.3$.
As the hidden domain wall state approaches either conduction band minima or valence band maxima, pinpointing its exact position becomes challenging.}
\end{figure}

The chirality inversion of the in-gap states of domain walls between the upper and lower gaps is more clearly seen if we plot energy spectra as a function of interchain coupling. %
Figure~\ref{fig.3.evolution3} shows the evolution of spectra of domain wall states with increasing interchain coupling via the single- to double-gap phase transition.
Regardless of the interchain coupling, two nontrivial domain wall states always exist in the lower gap.
On the other hand, additional domain wall states emerge when the upper gap opens.
Even though additional domain wall states are hidden in the region of $\alpha < \alpha_1$, the otherwise hidden domain wall states eventually emerge as the upper gap opens in the region of $\alpha_1 < \alpha < \alpha_2$.
In the region of $\alpha \ge \alpha_2$ where the upper gap is direct, the emerging domain wall states maintain their relative energy positions within the gap.
Therefore, each domain wall state's chirality (or the spectrum feature) is preserved during the evolution.

Before proceeding further, we briefly discuss the state 1 appearing in Figs.~\ref{fig.2.domain_wall} and \ref{fig.3.evolution3}.
The states labeled 1 are also possible hidden domain wall states between the first and second bands.
The LDOS maps in Fig.~\ref{fig.2.domain_wall}c,d clearly show the localized feature except for the AA-BA configuration in Fig.~\ref{fig.2.domain_wall}c.
In the AA-BA configuration, the domain wall state's spectrum lies too close to the top of the second band for given parameters, making it hard to discern between such a hidden domain wall state and the bulk state.
Moreover, the inversion symmetry protects the gap-closing between the first and second bands at the Brillouin zone boundary, as discussed in the previous subsection (see also Fig.~\ref{fig.1.gapphase}).
Therefore, the hidden domain wall states labeled 1 cannot manifest as in-gap states.
Consequently, we will focus on the other states from now on.
However, it is worth noting that the hidden domain wall states labeled 1 can indeed emerge as in-gap states by introducing a symmetry-breaking mechanism that opens up the gap in the system, as demonstrated in Supplementary Figure~1.

\textbf{Bulk-boundary correspondence.}
We now investigate the correspondence between the electronic states and the topological properties of the domain walls using Berry phase and Berry curvature via bulk-boundary correspondence~\cite{hasan2010, qi2011}.
In the context of one-dimensional systems, the electronic spectra of a domain wall state are related to the Berry phase difference between two groundstates that the domain wall interpolates~\cite{RM1982, qi2008topological}.

\begin{table}[b]
\caption{
\label{Tableberryphase}
\textbf{Berry phases up to the 2nd and 3rd bands for groundstates.}
The Berry phase is defined within a range of modulo $4\pi$ instead of the conventional $2\pi$ in accordance with the $Z_4$ property and the charge pumping phenomena occurring during the relevant adiabatic process, as discussed in the subsection Bulk-boundary correspondence.
}
{\renewcommand{\arraystretch}{1.3}%
\begin{tabular}{@{\extracolsep{8pt}} l  cccccc  }
\hline
\hline
 & ~~~AA~~~ & ~~~BA~~~ & ~~~BB~~~ & ~AB~~ & \\
\hline 
up to the 3rd Band & 0 & $\pi/2$ & $\pi$ & $3\pi/2$ &\\
up to the 2nd Band & 0 & $-\pi$ & $-2\pi$ & $-3\pi$ &\\
\hline
\hline
\end{tabular}
}
\end{table}

\begin{figure}[t]
\centering
\includegraphics[width=.8\linewidth]{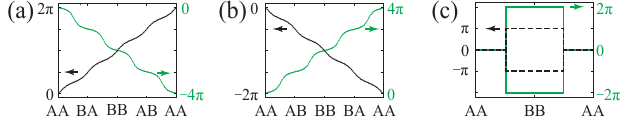}
\caption{\label{fig.4.cyclic_adiabatic}
\textbf{Evolution of Berry phase under cyclic adiabatic processes.} 
Evolution of Berry phase for \textbf{a} right-chiral, \textbf{b} left-chiral, and \textbf{c} achiral domain walls under the corresponding cyclic adiabatic processes.
These processes are generated as the corresponding chiral domain walls move from right to left along the one-dimensional chain. Black (green) color denotes the total Berry phase up to the third (second) bands from the lowest band.
Here, $(\alpha/t, \delta/t) = (0.7,0.7)$.
}
\end{figure}

Table~\ref{Tableberryphase} shows the calculated Berry phases for the four groundstates up to the second and third bands from the lowest one.
The well-separated electronic bands depicted in Fig.~\ref{fig.1.gapphase}c--e enable a band-by-band definition of the Berry phase, facilitating a more precise analysis of the system's electronic properties and their topological aspects.
All Berry phases are quantized as integer multiples of $\pi/2$, due to the $Z_4$ symmetry of the system~\cite{cheon2015, han2020} (The mathematical details are provided in Supplementary Notes 2 and 3).
The Berry phases up to the second band decrease as $0, -\pi, -2\pi$, and $-3\pi$ for AA, BA, BB, and AB groundstates. 
On the other hand, the Berry phases to the third band increase as $0, \pi/2, \pi$, and $3\pi/2$ for AA, BA, BB, and AB groundstates.
In contrast to the SSH model, some Berry phases are larger than $2\pi$.
It is noteworthy that the Berry phase is defined within a range of modulo $4\pi$ instead of the conventional $2\pi$ reflecting the evolution of the Wannier center and the $Z_4$ symmetry of the system~\cite{cheon2015, han2020}.
Such adjustment also accommodates the charge pumping phenomena occurring during the relevant adiabatic process, as discussed below. 

To clarify such Berry phases, as shown in Fig.~\ref{fig.4.cyclic_adiabatic}, we plot the continuous change of the Berry phase (or the evolution of the Wannier charge center~\cite{marzari1997, han2020}) up to the second and third bands for three types of cyclic adiabatic processes.
The three types of cyclic adiabatic processes are generated as the same types of chiral domain walls move from right to left along the one-dimensional chain.
Because there are three types of  chiral domain walls, we can identify three distinct cyclic adiabatic processes as follows:
(1) RC adiabatic process: AA$\rightarrow$BA$\rightarrow$BB$\rightarrow$AB$\rightarrow$AA, (2)
LC adiabatic process: AA$\rightarrow$AB$\rightarrow$BB$\rightarrow$BA$\rightarrow$AA,  and (3) AC adiabatic process: AA$\rightarrow$BB$\rightarrow$AA.

Above all, we will discuss the Berry phase under RC and LC adiabatic processes.
During the RC adiabatic process (Fig.~\ref{fig.4.cyclic_adiabatic}a), the total Berry phase up to the second band evolves from 0 to $2\pi$ while the one up to the third band changes from $0$ to $-4\pi$.
Conversely, the LC adiabatic process shows the opposite change of the Berry phase (Fig.~\ref{fig.4.cyclic_adiabatic}b).
As a result, two intriguing observations emerge. First, within a given adiabatic process type, the tendencies for the Berry phase up to the second band and the Berry phase up to the third band are diametrically opposite. Second, if we hold the chemical potential fixed, thereby keeping the band occupation constant, the variation tendencies of the Berry phases between the RC and LC adiabatic processes are also in opposition.
Such findings provide compelling topological evidence for the inversion of chirality between the upper and lower gaps, as well as between RC and LC chiral domain wall states. From a physical perspective, this opposite variation in the Berry phase is tantamount to the counter-directional charge pumping observed during the adiabatic process, which will be elaborated upon in the subsequent paragraph.

\begin{figure}[t]
\centering
\includegraphics[width=.7\linewidth]{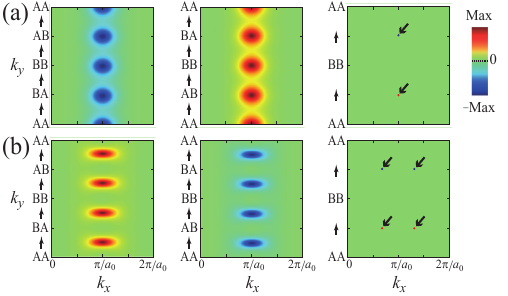}
\caption{\label{fig.5.berryphase}
\textbf{Berry curvature distributions.} 
Berry curvature distributions up to \textbf{a} third and \textbf{b} second bands under a cyclic adiabatic evolution using extended 2D Hamiltonians, normalized by the maximum absolute magnitude.
The color bar indicates the intensity of the normalized Berry curvature.
The cyclic adiabatic process is represented in terms of momentum $k_y$ through the dimensional extension~\cite{cheon2015,han2020}, enabling the calculation of Berry curvature within the two-dimensional Brillouin zone using extended 2D Hamiltonians.
The black arrows indicate singular points.
Here, $(\alpha/t, \delta/t) = (0.7,0.7)$.
}
\end{figure}

From the viewpoint of topology, the chirality of the in-gap state of a domain wall is determined by the direction of topological charge pumping under the adiabatic evolution from one groundstate to the other groundstate when the two groundstates are interpolated by the domain wall~\cite{rice1982, thouless1983, cheon2015, han2020, oh2021}. 
If the direction of the topological charge pumping is negative (positive), the electronic states are located below (above) the midgap.
Because the local information of such topological charge pumping is encoded in the Berry curvature during the adiabatic process, we can also identify the chirality of the in-gap state of a domain wall using Berry curvature distribution.
Notice that the direction of charge pumping (or the moving direction of the Wannier charge center) and the sign of Berry curvature are opposite. See more details in Supplementary Note 2.

First, let us consider the in-gap states in the lower gap.
For a right-chiral (left-chiral) domain wall connecting from AA to BA (from AA to AB) groundstates, the sign of Berry curvature distribution up to the second band from the lowest one, as shown in the first (second) panel of Fig.~\ref{fig.5.berryphase}b, is positive (negative), and hence, the in-gap states in the lower gap are located below (above) the midgap, as shown in Fig.~\ref{fig.3.evolution3}a,b.
Thus, the corresponding charge pumping under the RC and LC adiabatic processes are also opposite, as shown in Supplementary Figure~2.

Next, let us consider the in-gap states in the upper gap.
In this case, the directions of topological pumping are opposite compared to the lower gap case (Supplementary Figure~2),
which is consistent with the chirality inversion of chiral domain walls between upper and lower gaps.
Thus, for a right-chiral (left-chiral) domain wall connecting from AA to BA (from AA to AB), the sign of the Berry curvature distribution up to the third band in the first (second) panels of Fig.~\ref{fig.5.berryphase}a is negative (positive), and hence the in-gap states in the upper gap are located above (below) the midgap as shown in Fig.~\ref{fig.3.evolution3}a,b.

Finally, for the in-gap states of achiral domain walls in both upper and lower gaps,
there is no charge pumping because of the zero total Berry curvature during the adiabatic process (AA$\rightarrow$BB$\rightarrow$AA) as shown in the third panels of Fig.~\ref{fig.5.berryphase}a,b, which is consistent with the change of the Berry phase in Fig.~\ref{fig.4.cyclic_adiabatic}c.
This gives the symmetrically located electronic states, as indicated by the purple lines in Fig.~\ref{fig.3.evolution3}c.

\begin{figure}[t]
\centering
\includegraphics[width=.7\linewidth]{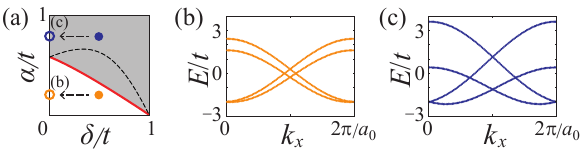}
\caption{\label{fig.6.without_dimerization}
\textbf{Band structures without horizontal dimerization.} 
\textbf{a} The same phase diagram in Fig.~\ref{fig.1.gapphase}\textbf{b}.
\textbf{b,c} Band structures for open circles in \textbf{a}: \textbf{b} for the orange circle and \textbf{c} for the blue circle.
In the absence of dimerization, the formation of the four Dirac points is the same for single- and double-gap phases as shown in \textbf{b} and \textbf{c}.
The parameters ($\alpha/t$, $\delta/t$) are (0.2, 0.0) for \textbf{b} and (0.8, 0.0) for \textbf{c}.
}
\end{figure}

Note that the same (opposite) Berry curvature distribution is repeated during the chiral (achiral) adiabatic process due to the system's $Z_4$ symmetry~\cite{cheon2015, han2020} as shown in Fig.~\ref{fig.5.berryphase}, which leads to the quantized Berry phases of the groundstates.
Furthermore, we find that the quantized Berry phases for the four groundstates in Table~\ref{Tableberryphase} are independent of the interchain coupling and dimerization strength.
This strongly implies that both the hidden chiral domain wall state and the in-gap chiral domain wall state possess consistent topological properties independent of the interchain coupling and dimerization.
Usually, a topological phase transition occurs when the gap between two energy bands closes and reopens with the change of topological invariants. 
Surprisingly, our study does not observe such a topological transition even after the upper gap opens, as the topological invariants remain unchanged. Note that the structure of Dirac points remains constant regardless of the interchain coupling in the absence of dimerization, as shown in Fig.~\ref{fig.6.without_dimerization}. This observation highlights the robustness of the topological properties in the double-chain model, independent of the interchain coupling.

\section*{Conclusion}
In summary, we have studied the emergence of the hidden topological domain wall states via gap engineering without requiring any topological phase transition
using a representative double-chain SSH model, specifically designed to break the chiral symmetry.
By varying the dimerization and interchain coupling, we constructed the phase diagram composed of single- and double-gap phases.
For a small interchain coupling, we found the chiral domain wall states with two localized in-gap states in the single gap.
However, for a larger interchain coupling, hidden domain wall states emerge, featuring only a single localized state in an additional gap. Intriguingly, the chirality of these emergent domain wall states in the second gap was found to be opposite compared to that of the original domain wall states in the first gap.
We validated this chirality inversion through spectrum inversion of the domain wall state and the observation of opposite charge pumping during the adiabatic process observed as the domain wall state moved slowly. The topological confirmation of the chirality of hidden chiral domain wall states was further supported through the analysis of the Berry curvature distribution.

This kind of chirality emergence is plentiful in nature and has many applications such as chirality-dependent light-matter devices~\cite{lininger2022}, chiral quantum optics~\cite{lodahl2017}, and chiral magnetic domain memory devices~\cite{dor2013, ryu2013}.
Therefore, we expect our theoretical approach can be applied to diverse topological systems such as In/Si(111)~\cite{kim2012, cheon2015}, Cl vacancies on Cl/Cu(100)~\cite{huda2020, jeong2022}, and photonic lattices as well as topological laser systems~\cite{st2017, bandres2018}.
For instance, our model system can be used as a multi-digit topological information carrier~\cite{kim2017} by engineering the gap structure and Fermi level.  
We also expect a new type of multi-frequency topological laser, where the topological single-mode lasing frequency can be selectively controlled.

\section*{Methods}
The band structures and phase diagrams in Figs.~\ref{fig.1.gapphase} and \ref{fig.6.without_dimerization} were studied using the Bloch Hamiltonian of the DC model, which is given by
\begin{eqnarray*}
\begin{scriptsize}
    H(\Delta^{(1)}, \Delta^{(2)}, k_x) = 
    \begin{pmatrix}
        0 & t^{(1)}_{+}e{^{i k_x \frac{a_0}{2}}}+t^{(1)}_{-}e{^{-i k_x \frac{a_0}{2}}} & \alpha e^{- i k_x \frac{a_0}{4}} & \alpha e^{i k_x \frac{a_0}{4}} \\
        t^{(1)}_{+}e{^{-i k_x \frac{a_0}{2}}}+t^{(1)}_{-}e{^{i k_x \frac{a_0}{2}}} & 0 & \alpha e^{ i k_x \frac{a_0}{4}} & \alpha e^{-i k_x \frac{a_0}{4}} \\
        \alpha e^{ i k_x \frac{a_0}{4}} & \alpha e^{-i k_x \frac{a_0}{4}} & 0 & t^{(2)}_{+}e{^{i k_x \frac{a_0}{2}}}+t^{(2)}_{-}e{^{-i k_x \frac{a_0}{2}}} \\
        \alpha e^{-i k_x \frac{a_0}{4}} & \alpha e^{i k_x \frac{a_0}{4}} & t^{(2)}_{+}e{^{-i k_x \frac{a_0}{2}}}+t^{(2)}_{-}e{^{i k_x \frac{a_0}{2}}} & 0 \\
    \end{pmatrix},
\end{scriptsize}
\end{eqnarray*}
where $t^{(i)}_{\pm} = t \pm \Delta^{(i)}$ with energy-valued dimerization $\Delta^{(i)}$ for the $i$-th chain.

To obtain the spectra and LDOS for the RC, LC, and AC chiral domain walls states in Figs.~\ref{fig.2.domain_wall} and \ref{fig.3.evolution3}, we used the tight-binding method for the finite system having $4n+3$, $4n+1$, and $4n+2$ atoms with $n=200$, respectively.
Such boundary conditions ensure the absence of localized edge states at both ends.
The dimerization patterns for the domain wall states were simulated using the position-dependent dimerizations and hyperbolic tangent functions: 
$ \Delta^{(i)}(x) = \pm \delta \tanh{(x/ \xi)}$, with $\xi$ being the characteristic width of the domain wall, where $\xi = 1.5 a_0$ in Figs.~\ref{fig.2.domain_wall} and \ref{fig.3.evolution3}.

For the Berry phase and Berry curvature in Figs.~\ref{fig.4.cyclic_adiabatic} and \ref{fig.5.berryphase}, we took into account a cyclic adiabatic process of a 1D Hamiltonian, denoted as $H(k_x, \tau)$, where $\tau$ is time for the adiabatic evolution. By extending the 1D lattice system into a 2D lattice system, we replace the time evolution with momentum $k_y$ in an extra dimension.
Then, we constructed the 2D Hamiltonian $H_{\text{2D}}(k_x, k_y)$
such that $H_{\text{2D}}(k_x, k_y=0) = H(k_x, \tau=0)$ and $H_{\text{2D}}(k_x, k_y=2 \pi) = H(k_x, \tau=T)$, where $T$ represents the period of  the corresponding cyclic adiabatic process.
Using this 2D Hamiltonian, we have calculated the Berry phase and Berry curvature.
Further comprehensive information can be found in Supplementary Notes 2--4.

\section*{Data availability}
The data used in this paper are available from T.-H.K. or S.C. on reasonable request.

\section*{Acknowledgments}
This work was supported by the National Research Foundation of Korea (NRF) funded by the Ministry of Science and ICT (MSIT), South Korea (Grants No. NRF-2021R1H1A1013517, NRF-2022R1A2C1011646, NRF-2022M3H3A1085772, NRF-2021R1A6A1A10042944, and 2022M3H4A1A04074153).
This work was also supported by Quantum Simulator Development Project for Materials Innovation through the NRF funded by the MSIT, South Korea (Grant No. NRF-2023M3K5A1094813).
S.-H.H. and S.C. acknowledge support from the POSCO Science Fellowship of POSCO TJ Park Foundation.

\section*{Author contributions}
S.-G.J., T.-H.K., and S.C. conceived and designed the project.
S.-G.J. and S.-H.H. performed the tight-binding calculations and analyzed the results under the supervision of T.-H.K. and S.C.
All the authors discussed the results and contributed to the writing of the manuscript.

\section*{Competing interests}
The authors declare no competing interests.

\section*{Additional information}

\textbf{Supplementary information} The online version contains supplementary information available at https://doi.org/10.1038/xxxx.

\textbf{Correspondence} and requests for materials should be addressed to Tae-Hwan Kim or Sangmo Cheon.


\begin{thebibliography}{10}
\expandafter\ifx\csname url\endcsname\relax
  \def\url#1{\texttt{#1}}\fi
\expandafter\ifx\csname urlprefix\endcsname\relax\def\urlprefix{URL }\fi
\providecommand{\bibinfo}[2]{#2}
\providecommand{\eprint}[2][]{\url{#2}}

\bibitem{hasan2010}
\bibinfo{author}{Hasan, M.~Z.} \& \bibinfo{author}{Kane, C.~L.}
\newblock \bibinfo{title}{\textit{{Colloquium}} : {Topological} insulators}.
\newblock \emph{\bibinfo{journal}{Rev. Mod. Phys.}}
  \textbf{\bibinfo{volume}{82}}, \bibinfo{pages}{3045--3067}
  (\bibinfo{year}{2010}).

\bibitem{qi2011}
\bibinfo{author}{Qi, X.-L.} \& \bibinfo{author}{Zhang, S.-C.}
\newblock \bibinfo{title}{Topological insulators and superconductors}.
\newblock \emph{\bibinfo{journal}{Rev. Mod. Phys.}}
  \textbf{\bibinfo{volume}{83}}, \bibinfo{pages}{1057--1110}
  (\bibinfo{year}{2011}).

\bibitem{kitaev2001}
\bibinfo{author}{Kitaev, A.~Y.}
\newblock \bibinfo{title}{Unpaired {Majorana} fermions in quantum wires}.
\newblock \emph{\bibinfo{journal}{Phys.-Usp.}} \textbf{\bibinfo{volume}{44}},
  \bibinfo{pages}{131--136} (\bibinfo{year}{2001}).

\bibitem{elliott2015}
\bibinfo{author}{Elliott, S.~R.} \& \bibinfo{author}{Franz, M.}
\newblock \bibinfo{title}{\textit{{Colloquium}} : {Majorana} fermions in
  nuclear, particle, and solid-state physics}.
\newblock \emph{\bibinfo{journal}{Rev. Mod. Phys.}}
  \textbf{\bibinfo{volume}{87}}, \bibinfo{pages}{137--163}
  (\bibinfo{year}{2015}).

\bibitem{xu2020}
\bibinfo{author}{Xu, S.-Y.} \emph{et~al.}
\newblock \bibinfo{title}{Spontaneous gyrotropic electronic order in a
  transition-metal dichalcogenide}.
\newblock \emph{\bibinfo{journal}{Nature}} \textbf{\bibinfo{volume}{578}},
  \bibinfo{pages}{545--549} (\bibinfo{year}{2020}).

\bibitem{jiang2021}
\bibinfo{author}{Jiang, Y.-X.} \emph{et~al.}
\newblock \bibinfo{title}{Unconventional chiral charge order in kagome
  superconductor {KV$_3$Sb$_5$}}.
\newblock \emph{\bibinfo{journal}{Nat. Mater.}} \textbf{\bibinfo{volume}{20}},
  \bibinfo{pages}{1353--1357} (\bibinfo{year}{2021}).

\bibitem{kim2022}
\bibinfo{author}{Kim, S.-W.}, \bibinfo{author}{Kim, H.-J.},
  \bibinfo{author}{Cheon, S.} \& \bibinfo{author}{Kim, T.-H.}
\newblock \bibinfo{title}{Circular {Dichroism} of {Emergent} {Chiral}
  {Stacking} {Orders} in {Quasi}-{One}-{Dimensional} {Charge} {Density}
  {Waves}}.
\newblock \emph{\bibinfo{journal}{Phys. Rev. Lett.}}
  \textbf{\bibinfo{volume}{128}}, \bibinfo{pages}{046401}
  (\bibinfo{year}{2022}).

\bibitem{st2017}
\bibinfo{author}{St-Jean, P.} \emph{et~al.}
\newblock \bibinfo{title}{Lasing in topological edge states of a
  one-dimensional lattice}.
\newblock \emph{\bibinfo{journal}{Nat. Photon.}} \textbf{\bibinfo{volume}{11}},
  \bibinfo{pages}{651--656} (\bibinfo{year}{2017}).

\bibitem{bandres2018}
\bibinfo{author}{Bandres, M.~A.} \emph{et~al.}
\newblock \bibinfo{title}{Topological insulator laser: Experiments}.
\newblock \emph{\bibinfo{journal}{Science}} \textbf{\bibinfo{volume}{359}},
  \bibinfo{pages}{1231} (\bibinfo{year}{2018}).

\bibitem{SSH1979}
\bibinfo{author}{Su, W.~P.}, \bibinfo{author}{Schrieffer, J.~R.} \&
  \bibinfo{author}{Heeger, A.~J.}
\newblock \bibinfo{title}{Solitons in {Polyacetylene}}.
\newblock \emph{\bibinfo{journal}{Phys. Rev. Lett.}}
  \textbf{\bibinfo{volume}{42}}, \bibinfo{pages}{1698--1701}
  (\bibinfo{year}{1979}).

\bibitem{RM1982}
\bibinfo{author}{Rice, M.~J.} \& \bibinfo{author}{Mele, E.~J.}
\newblock \bibinfo{title}{Elementary {Excitations} of a {Linearly} {Conjugated}
  {Diatomic} {Polymer}}.
\newblock \emph{\bibinfo{journal}{Phys. Rev. Lett.}}
  \textbf{\bibinfo{volume}{49}}, \bibinfo{pages}{1455--1459}
  (\bibinfo{year}{1982}).

\bibitem{jackiw1976}
\bibinfo{author}{Jackiw, R.} \& \bibinfo{author}{Rebbi, C.}
\newblock \bibinfo{title}{Solitons with fermion number \textonehalf}.
\newblock \emph{\bibinfo{journal}{Phys. Rev. D}} \textbf{\bibinfo{volume}{13}},
  \bibinfo{pages}{3398--3409} (\bibinfo{year}{1976}).

\bibitem{goldstone1981}
\bibinfo{author}{Goldstone, J.} \& \bibinfo{author}{Wilczek, F.}
\newblock \bibinfo{title}{Fractional {Quantum} {Numbers} on {Solitons}}.
\newblock \emph{\bibinfo{journal}{Phys. Rev. Lett.}}
  \textbf{\bibinfo{volume}{47}}, \bibinfo{pages}{986--989}
  (\bibinfo{year}{1981}).

\bibitem{jackiw1981}
\bibinfo{author}{Jackiw, R.} \& \bibinfo{author}{Schrieffer, J.~R.}
\newblock \bibinfo{title}{Solitons with fermion number $\frac{1}{2}$ in
  condensed matter and relativistic field theories}.
\newblock \emph{\bibinfo{journal}{Nucl. Phys. B}}
  \textbf{\bibinfo{volume}{190}}, \bibinfo{pages}{253--265}
  (\bibinfo{year}{1981}).

\bibitem{han2020}
\bibinfo{author}{Han, S.-H.}, \bibinfo{author}{Jeong, S.-G.},
  \bibinfo{author}{Kim, S.-W.}, \bibinfo{author}{Kim, T.-H.} \&
  \bibinfo{author}{Cheon, S.}
\newblock \bibinfo{title}{Topological features of ground states and topological
  solitons in generalized {Su}-{Schrieffer}-{Heeger} models using generalized
  time-reversal, particle-hole, and chiral symmetries}.
\newblock \emph{\bibinfo{journal}{Phys. Rev. B}}
  \textbf{\bibinfo{volume}{102}}, \bibinfo{pages}{235411}
  (\bibinfo{year}{2020}).

\bibitem{heeger1988}
\bibinfo{author}{Heeger, A.~J.}, \bibinfo{author}{Kivelson, S.},
  \bibinfo{author}{Schrieffer, J.} \& \bibinfo{author}{Su, W.-P.}
\newblock \bibinfo{title}{Solitons in conducting polymers}.
\newblock \emph{\bibinfo{journal}{Rev. Mod. Phys.}}
  \textbf{\bibinfo{volume}{60}}, \bibinfo{pages}{781--850}
  (\bibinfo{year}{1988}).

\bibitem{atala2013}
\bibinfo{author}{Atala, M.} \emph{et~al.}
\newblock \bibinfo{title}{Direct measurement of the {Zak} phase in topological
  {Bloch} bands}.
\newblock \emph{\bibinfo{journal}{Nat. Phys.}} \textbf{\bibinfo{volume}{9}},
  \bibinfo{pages}{795--800} (\bibinfo{year}{2013}).

\bibitem{cooper2019}
\bibinfo{author}{Cooper, N.~R.}, \bibinfo{author}{Dalibard, J.} \&
  \bibinfo{author}{Spielman, I.~B.}
\newblock \bibinfo{title}{Topological bands for ultracold atoms}.
\newblock \emph{\bibinfo{journal}{Rev. Mod. Phys.}}
  \textbf{\bibinfo{volume}{91}}, \bibinfo{pages}{015005}
  (\bibinfo{year}{2019}).

\bibitem{drost2017}
\bibinfo{author}{Drost, R.}, \bibinfo{author}{Ojanen, T.},
  \bibinfo{author}{Harju, A.} \& \bibinfo{author}{Liljeroth, P.}
\newblock \bibinfo{title}{Topological states in engineered atomic lattices}.
\newblock \emph{\bibinfo{journal}{Nat. Phys.}} \textbf{\bibinfo{volume}{13}},
  \bibinfo{pages}{668--671} (\bibinfo{year}{2017}).

\bibitem{huda2020}
\bibinfo{author}{Huda, M.~N.}, \bibinfo{author}{Kezilebieke, S.},
  \bibinfo{author}{Ojanen, T.}, \bibinfo{author}{Drost, R.} \&
  \bibinfo{author}{Liljeroth, P.}
\newblock \bibinfo{title}{Tuneable topological domain wall states in engineered
  atomic chains}.
\newblock \emph{\bibinfo{journal}{npj Quantum Mater.}}
  \textbf{\bibinfo{volume}{5}}, \bibinfo{pages}{17} (\bibinfo{year}{2020}).

\bibitem{meier2016}
\bibinfo{author}{Meier, E.~J.}, \bibinfo{author}{An, F.~A.} \&
  \bibinfo{author}{Gadway, B.}
\newblock \bibinfo{title}{Observation of the topological soliton state in the
  {Su}--{Schrieffer}--{Heeger} model}.
\newblock \emph{\bibinfo{journal}{Nat. Commun.}} \textbf{\bibinfo{volume}{7}},
  \bibinfo{pages}{13986} (\bibinfo{year}{2016}).

\bibitem{ozawa2019}
\bibinfo{author}{Ozawa, T.} \emph{et~al.}
\newblock \bibinfo{title}{Topological photonics}.
\newblock \emph{\bibinfo{journal}{Rev. Mod. Phys.}}
  \textbf{\bibinfo{volume}{91}}, \bibinfo{pages}{015006}
  (\bibinfo{year}{2019}).

\bibitem{zhou2017}
\bibinfo{author}{Zhou, X.-F.} \emph{et~al.}
\newblock \bibinfo{title}{Dynamically {Manipulating} {Topological} {Physics}
  and {Edge} {Modes} in a {Single} {Degenerate} {Optical} {Cavity}}.
\newblock \emph{\bibinfo{journal}{Phys. Rev. Lett.}}
  \textbf{\bibinfo{volume}{118}}, \bibinfo{pages}{083603}
  (\bibinfo{year}{2017}).

\bibitem{zeng2021}
\bibinfo{author}{Zeng, L.-S.}, \bibinfo{author}{Shen, Y.-X.},
  \bibinfo{author}{Peng, Y.-G.}, \bibinfo{author}{Zhao, D.-G.} \&
  \bibinfo{author}{Zhu, X.-F.}
\newblock \bibinfo{title}{Selective {Topological} {Pumping} for {Robust},
  {Efficient}, and {Asymmetric} {Sound} {Energy} {Transfer} in a {Dynamically}
  {Coupled} {Cavity} {Chain}}.
\newblock \emph{\bibinfo{journal}{Phys. Rev. Appl.}}
  \textbf{\bibinfo{volume}{15}}, \bibinfo{pages}{064018}
  (\bibinfo{year}{2021}).

\bibitem{berry1984}
\bibinfo{author}{Berry, M.~V.}
\newblock \bibinfo{title}{Quantal phase factors accompanying adiabatic
  changes}.
\newblock \emph{\bibinfo{journal}{Proc. R. Soc. London, Ser. A}}
  \textbf{\bibinfo{volume}{392}}, \bibinfo{pages}{45--57}
  (\bibinfo{year}{1984}).

\bibitem{hatsugai1993}
\bibinfo{author}{Hatsugai, Y.}
\newblock \bibinfo{title}{Chern number and edge states in the integer quantum
  {Hall} effect}.
\newblock \emph{\bibinfo{journal}{Phys. Rev. Lett.}}
  \textbf{\bibinfo{volume}{71}}, \bibinfo{pages}{3697--3700}
  (\bibinfo{year}{1993}).

\bibitem{bernevig2013topological}
\bibinfo{author}{Bernevig, B.} \& \bibinfo{author}{Hughes, T.}
\newblock \emph{\bibinfo{title}{Topological Insulators and Topological
  Superconductors}} (\bibinfo{publisher}{Princeton University Press},
  \bibinfo{year}{2013}).

\bibitem{arkinstall2017}
\bibinfo{author}{Arkinstall, J.}, \bibinfo{author}{Teimourpour, M.~H.},
  \bibinfo{author}{Feng, L.}, \bibinfo{author}{El-Ganainy, R.} \&
  \bibinfo{author}{Schomerus, H.}
\newblock \bibinfo{title}{Topological tight-binding models from nontrivial
  square roots}.
\newblock \emph{\bibinfo{journal}{Phys. Rev. B}} \textbf{\bibinfo{volume}{95}},
  \bibinfo{pages}{165109} (\bibinfo{year}{2017}).

\bibitem{Zurita2021}
\bibinfo{author}{Zurita, J.}, \bibinfo{author}{Creffield, C.} \&
  \bibinfo{author}{Platero, G.}
\newblock \bibinfo{title}{Tunable zero modes and quantum interferences in
  flat-band topological insulators}.
\newblock \emph{\bibinfo{journal}{Quantum}} \textbf{\bibinfo{volume}{5}},
  \bibinfo{pages}{591} (\bibinfo{year}{2021}).

\bibitem{Luo2022}
\bibinfo{author}{Luo, T.}, \bibinfo{author}{Guan, X.}, \bibinfo{author}{Fan,
  J.}, \bibinfo{author}{Chen, G.} \& \bibinfo{author}{Jia, S.-T.}
\newblock \bibinfo{title}{Topological phases and type-{{II}} edge state in
  two-leg-coupled {{Su}}\textendash{{Schrieffer}}\textendash{{Heeger}} chains}.
\newblock \emph{\bibinfo{journal}{Chinese Phys. B}}
  \textbf{\bibinfo{volume}{31}}, \bibinfo{pages}{014208}
  (\bibinfo{year}{2022}).

\bibitem{Matveeva2023}
\bibinfo{author}{Matveeva, P.} \emph{et~al.}
\newblock \bibinfo{title}{One-dimensional noninteracting topological insulators
  with chiral symmetry}.
\newblock \emph{\bibinfo{journal}{Phys. Rev. B}}
  \textbf{\bibinfo{volume}{107}}, \bibinfo{pages}{075422}
  (\bibinfo{year}{2023}).

\bibitem{cheon2015}
\bibinfo{author}{Cheon, S.}, \bibinfo{author}{Kim, T.-H.},
  \bibinfo{author}{Lee, S.-H.} \& \bibinfo{author}{Yeom, H.~W.}
\newblock \bibinfo{title}{Chiral solitons in a coupled double {Peierls} chain}.
\newblock \emph{\bibinfo{journal}{Science}} \textbf{\bibinfo{volume}{350}},
  \bibinfo{pages}{182--185} (\bibinfo{year}{2015}).

\bibitem{jeong2022}
\bibinfo{author}{Jeong, S.-G.} \& \bibinfo{author}{Kim, T.-H.}
\newblock \bibinfo{title}{Topological and trivial domain wall states in
  engineered atomic chains}.
\newblock \emph{\bibinfo{journal}{npj Quantum Mater.}}
  \textbf{\bibinfo{volume}{7}}, \bibinfo{pages}{22} (\bibinfo{year}{2022}).

\bibitem{oh2021}
\bibinfo{author}{Oh, C.-g.}, \bibinfo{author}{Han, S.-H.},
  \bibinfo{author}{Jeong, S.-G.}, \bibinfo{author}{Kim, T.-H.} \&
  \bibinfo{author}{Cheon, S.}
\newblock \bibinfo{title}{Particle-antiparticle duality and fractionalization
  of topological chiral solitons}.
\newblock \emph{\bibinfo{journal}{Sci. Rep.}} \textbf{\bibinfo{volume}{11}},
  \bibinfo{pages}{1013} (\bibinfo{year}{2021}).

\bibitem{schnyder2008}
\bibinfo{author}{Schnyder, A.~P.}, \bibinfo{author}{Ryu, S.},
  \bibinfo{author}{Furusaki, A.} \& \bibinfo{author}{Ludwig, A. W.~W.}
\newblock \bibinfo{title}{Classification of topological insulators and
  superconductors in three spatial dimensions}.
\newblock \emph{\bibinfo{journal}{Phys. Rev. B}} \textbf{\bibinfo{volume}{78}},
  \bibinfo{pages}{195125} (\bibinfo{year}{2008}).

\bibitem{chiu2016}
\bibinfo{author}{Chiu, C.-K.}, \bibinfo{author}{Teo, J. C.~Y.},
  \bibinfo{author}{Schnyder, A.~P.} \& \bibinfo{author}{Ryu, S.}
\newblock \bibinfo{title}{Classification of topological quantum matter with
  symmetries}.
\newblock \emph{\bibinfo{journal}{Rev. Mod. Phys.}}
  \textbf{\bibinfo{volume}{88}}, \bibinfo{pages}{035005}
  (\bibinfo{year}{2016}).

\bibitem{SSH1980}
\bibinfo{author}{Su, W.~P.}, \bibinfo{author}{Schrieffer, J.~R.} \&
  \bibinfo{author}{Heeger, A.~J.}
\newblock \bibinfo{title}{Soliton excitations in polyacetylene}.
\newblock \emph{\bibinfo{journal}{Phys. Rev. B}} \textbf{\bibinfo{volume}{22}},
  \bibinfo{pages}{2099--2111} (\bibinfo{year}{1980}).

\bibitem{SSH1988}
\bibinfo{author}{Heeger, A.~J.}, \bibinfo{author}{Kivelson, S.},
  \bibinfo{author}{Schrieffer, J.~R.} \& \bibinfo{author}{Su, W.~P.}
\newblock \bibinfo{title}{Solitons in conducting polymers}.
\newblock \emph{\bibinfo{journal}{Rev. Mod. Phys.}}
  \textbf{\bibinfo{volume}{60}}, \bibinfo{pages}{781--850}
  (\bibinfo{year}{1988}).

\bibitem{kim2017}
\bibinfo{author}{Kim, T.-H.}, \bibinfo{author}{Cheon, S.} \&
  \bibinfo{author}{Yeom, H.~W.}
\newblock \bibinfo{title}{Switching chiral solitons for algebraic operation of
  topological quaternary digits}.
\newblock \emph{\bibinfo{journal}{Nat. Phys.}} \textbf{\bibinfo{volume}{13}},
  \bibinfo{pages}{444--447} (\bibinfo{year}{2017}).

\bibitem{qi2008topological}
\bibinfo{author}{Qi, X.-L.}, \bibinfo{author}{Hughes, T.~L.} \&
  \bibinfo{author}{Zhang, S.-C.}
\newblock \bibinfo{title}{Topological field theory of time-reversal invariant
  insulators}.
\newblock \emph{\bibinfo{journal}{Phys. Rev. B}} \textbf{\bibinfo{volume}{78}},
  \bibinfo{pages}{195424} (\bibinfo{year}{2008}).

\bibitem{marzari1997}
\bibinfo{author}{Marzari, N.} \& \bibinfo{author}{Vanderbilt, D.}
\newblock \bibinfo{title}{Maximally localized generalized {Wannier} functions
  for composite energy bands}.
\newblock \emph{\bibinfo{journal}{Phys. Rev. B}} \textbf{\bibinfo{volume}{56}},
  \bibinfo{pages}{12847--12865} (\bibinfo{year}{1997}).

\bibitem{rice1982}
\bibinfo{author}{Rice, M.~J.} \& \bibinfo{author}{Mele, E.~J.}
\newblock \bibinfo{title}{Elementary {Excitations} of a {Linearly} {Conjugated}
  {Diatomic} {Polymer}}.
\newblock \emph{\bibinfo{journal}{Phys. Rev. Lett.}}
  \textbf{\bibinfo{volume}{49}}, \bibinfo{pages}{1455--1459}
  (\bibinfo{year}{1982}).

\bibitem{thouless1983}
\bibinfo{author}{Thouless, D.~J.}
\newblock \bibinfo{title}{Quantization of particle transport}.
\newblock \emph{\bibinfo{journal}{Phys. Rev. B}} \textbf{\bibinfo{volume}{27}},
  \bibinfo{pages}{6083--6087} (\bibinfo{year}{1983}).

\bibitem{lininger2022}
\bibinfo{author}{Lininger, A.} \emph{et~al.}
\newblock \bibinfo{title}{Chirality in light–matter interaction}.
\newblock \emph{\bibinfo{journal}{Adv. Mater.}} \textbf{\bibinfo{volume}{N/A}},
  \bibinfo{pages}{2107325} (\bibinfo{year}{2022}).

\bibitem{lodahl2017}
\bibinfo{author}{Lodahl, P.} \emph{et~al.}
\newblock \bibinfo{title}{Chiral quantum optics}.
\newblock \emph{\bibinfo{journal}{Nature}} \textbf{\bibinfo{volume}{541}},
  \bibinfo{pages}{473--480} (\bibinfo{year}{2017}).

\bibitem{dor2013}
\bibinfo{author}{Dor, O.~B.}, \bibinfo{author}{Yochelis, S.},
  \bibinfo{author}{Mathew, S.~P.}, \bibinfo{author}{Naaman, R.} \&
  \bibinfo{author}{Paltiel, Y.}
\newblock \bibinfo{title}{A chiral-based magnetic memory device without a
  permanent magnet}.
\newblock \emph{\bibinfo{journal}{Nat. Commun.}} \textbf{\bibinfo{volume}{4}},
  \bibinfo{pages}{2256} (\bibinfo{year}{2013}).

\bibitem{ryu2013}
\bibinfo{author}{Ryu, K.-S.}, \bibinfo{author}{Thomas, L.},
  \bibinfo{author}{Yang, S.-H.} \& \bibinfo{author}{Parkin, S.}
\newblock \bibinfo{title}{Chiral spin torque at magnetic domain walls}.
\newblock \emph{\bibinfo{journal}{Nat. Nanotechnol.}}
  \textbf{\bibinfo{volume}{8}}, \bibinfo{pages}{527--533}
  (\bibinfo{year}{2013}).

\bibitem{kim2012}
\bibinfo{author}{Kim, T.-H.} \& \bibinfo{author}{Yeom, H.~W.}
\newblock \bibinfo{title}{Topological {Solitons} versus {Nonsolitonic} {Phase}
  {Defects} in a {Quasi}-{One}-{Dimensional} {Charge}-{Density} {Wave}}.
\newblock \emph{\bibinfo{journal}{Phys. Rev. Lett.}}
  \textbf{\bibinfo{volume}{109}}, \bibinfo{pages}{246802}
  (\bibinfo{year}{2012}).

\end{thebibliography}

\end{document}